%% file: HEApaperMixed.tex
\documentclass[%
 reprint,
superscriptaddress,
 amsmath,amssymb,
 aps,
pra,
]{revtex4-1}

\usepackage{graphicx}
\usepackage{dcolumn}
\usepackage{bm}
\usepackage{hyperref}
\usepackage[mathlines]{lineno}
\usepackage{xcolor}
\makeindex
\begin{document}

\preprint{APS/123-QED}

\title{ Multiscale nanoindentation
modeling of concentrated solid solutions: A continuum plasticity model}
\author{K. Frydrych}
\email[Corresponding author: ]{karol.frydrych@ncbj.gov.pl}

 \affiliation{NOMATEN Centre of Excellence, National Centre for Nuclear Research, ul. A. Sołtana 7, 05-400 Otwock, Poland}
 \affiliation{Institute of Fundamental Technological Research,
 Polish Academy of Sciences, Pawińskiego 5b, Warsaw, 02-106, Poland}
 \author{F. J. Dom\'inguez-Guti\'errez}
 \affiliation{NOMATEN Centre of Excellence, National Centre for Nuclear Research, ul. A. Sołtana 7, 05-400 Otwock, Poland}
 \author{M. Alava}
 \affiliation{NOMATEN Centre of Excellence, National Centre for Nuclear Research, ul. A. Sołtana 7, 05-400 Otwock, Poland}
  \affiliation{Department of Applied Physics, Aalto University, P.O. Box 11000, 00076 Aalto, Espoo, Finland}
  \author{S. Papanikolaou}
   \affiliation{NOMATEN Centre of Excellence, National Centre for Nuclear Research, ul. A. Sołtana 7, 05-400 Otwock, Poland}

\date{\today}

\begin{abstract}
Recently developed single-phase concentrated solid-solution 
alloys (CSAs) contain multiple elemental species in high 
concentrations with different elements randomly arranged
on a crystalline lattice. 
These chemically disordered materials present excellent physical
properties, including high-temperature thermal stability and 
hardness, with promising applications to industries at extreme
operating environments. 
The aim of this paper is to present a continuum plasticity model
accounting for the first time for the behaviour of a equiatomic
five-element CSA, that forms a face-centered cubic lattice. 
The inherent disorder associated with the lattice distortions 
caused by an almost equiatomic distribution of atoms, is 
captured by a single parameter $\alpha$ that quantifies the 
relative importance of an isotropic plastic contribution 
to the model. 
This results in multiple plasticity mechanisms that go beyond 
crystallographic symmetry-based ones, common in the case of 
conventional single element metals.
We perform molecular dynamics simulations of equiatomic CSAs: NiFe, 
NiFeCr, NiFeCrCo, and Cantor alloys to validate the proposed 
continuum model which is implemented in the finite element method
and applied to model nanoindentation tests for three different 
crystallographic orientations.
We obtain the representative volume element model by tracking the 
combined model yield surface. 
\end{abstract}

\maketitle
\input{Introduction/intro}
\input{MD/MD_method}
\input{Continuum/continuum}
\input{Results/results}
\input{MD/MD_results}
\input{Discussion/discussion}
\input{Conclusions/conclusion}

\section*{Acknowledgments}
We acknowledge support from the European Union Horizon 2020 research
 and innovation program under grant agreement no. 857470 and from the 
 European Regional Development Fund via the Foundation for Polish 
 Science International Research Agenda PLUS program grant 
 No. MAB PLUS/2018/8. We acknowledge the computational resources 
 provided by the High Performance Cluster at the National Centre 
 for Nuclear Research in Poland.
 Karol Frydrych acknowledges fruitful discussions with prof. Katarzyna Kowalczyk-Gajewska from IPPT PAN.
 
\appendix
\section{Derivation of the velocity gradient for the isotropic part of the model}
We build the isotropic plasticity model starting from the Perzyna-type viscoplastic model defined within the infinitesimal strain theory. In such a model, the J2 yield surface appears:
$$
\mathcal{F} = \sigma_{eq} - \sigma_y
$$
and the plastic part of the velocity framework is defined as follows:
$$
\hat{\mathbf{L}}^p=
\frac{\left\langle \mathcal{F} \right\rangle}{\eta}
\frac{\partial \mathcal{F}}{\partial \boldsymbol{\sigma}}.
$$
When we represent the tensors in the index notation, this has the following form:
$$
\hat{\mathbf{L}}^p=
\frac{\left\langle \mathcal{F} \right\rangle}{\eta}
\frac{\partial \sqrt{\frac{3}{2}s_{ij}s_{ij}}-\sigma_y}{\partial {\sigma_{kl}}}
=
\frac{\left\langle \mathcal{F} \right\rangle}{\eta \sqrt{\frac{3}{2}s_{ij}s_{ij}}}
\frac{1}{2}\frac{3}{2}\frac{\partial s_{ij}s_{ij}}{\partial {\sigma_{kl}}}
=$$
$$=
\frac{\left\langle \mathcal{F} \right\rangle}{\eta \sigma_{eq}}
\frac{3}{4}\frac{\partial \left( \sigma_{ij}\sigma_{ij} - \frac{2}{3} \sigma_{mm} \sigma_{ij}\delta_{ij} + \frac{1}{9}\sigma_{mm}\sigma_{mm}\delta_{ij}\delta_{ij} \right)}{\partial {\sigma_{kl}}}=
$$
$$=
\frac{\left\langle \mathcal{F} \right\rangle}{\eta \sigma_{eq}}
\frac{3}{4}\frac{\partial \left( \sigma_{ij}\sigma_{ij} - \frac{2}{3} \sigma_{mm} \sigma_{ij}\delta_{ij} + \frac{1}{3}\sigma_{mm}\sigma_{mm} \right)}{\partial {\sigma_{kl}}}
=$$
$$=
\frac{\left\langle \mathcal{F} \right\rangle}{\eta \sigma_{eq}}\frac{3}{4}
\left(
2\delta_{ik}\delta_{jl}\sigma_{ij} - \frac{2}{3}\delta_{kl}\delta_{ij}\sigma_{ij}
-\frac{2}{3}\sigma_{mm}\delta_{kl}
+\frac{2}{3}\delta_{kl}\sigma_{mm}
\right)
=
$$
$$=
\frac{\left\langle \mathcal{F} \right\rangle}{\eta \sigma_{eq}}\frac{3}{2}
\left(
\sigma_{kl}-\frac{1}{3}\sigma_{mm}\delta_{kl}
\right)=
\frac{\left\langle \mathcal{F} \right\rangle}{\eta \sigma_{eq}}\frac{3}{2}s_{kl}.
$$
Which, in direct notation is equivalent to:
$$
\hat{\mathbf{L}}^p= \frac{1}{\eta}
\frac{3}{2} \frac{\left\langle \mathcal{F} \right\rangle}{\sigma_{eq}}\mathbf{s}.
$$

On the other hand, when switching to the finite strain formulation the derivative is taken wrt. to the Kirchhoff stress $\boldsymbol{\tau} = J \boldsymbol{\sigma}$. 
Therefore:
$$
\hat{\mathbf{L}}^p=
\frac{\left\langle \mathcal{F} \right\rangle}{\eta}
\frac{\partial \sqrt{\frac{3}{2}s_{ij}s_{ij}}-\sigma_y}{\partial {\tau_{kl}}}
=
\frac{\left\langle \mathcal{F} \right\rangle}{\eta \sqrt{\frac{3}{2}s_{ij}s_{ij}}}
\frac{1}{2}\frac{3}{2}\frac{\partial s_{ij}s_{ij}}{\partial {\tau_{kl}}}
=$$
$$=
\frac{\left\langle \mathcal{F} \right\rangle}{\eta \sigma_{eq}}
\frac{3}{4}\frac{1}{J^2}\frac{\partial \left( \tau_{ij}\tau_{ij} - \frac{2}{3} \tau_{mm} \tau_{ij}\delta_{ij} + \frac{1}{9}\tau_{mm}\tau_{mm}\delta_{ij}\delta_{ij} \right)}{\partial {\tau_{kl}}}=
$$
$$=
\frac{\left\langle \mathcal{F} \right\rangle}{\eta \sigma_{eq}}
\frac{3}{4}\frac{1}{J^2}\frac{\partial \left( \tau_{ij}\tau_{ij} - \frac{2}{3} \tau_{mm} \tau_{ij}\delta_{ij} + \frac{1}{3}\tau_{mm}\tau_{mm} \right)}{\partial {\tau_{kl}}}
=$$
$$=
\frac{\left\langle \mathcal{F} \right\rangle}{\eta \sigma_{eq}}\frac{3}{4}\frac{1}{J^2}
\left(
2\delta_{ik}\delta_{jl}\tau_{ij} - \frac{2}{3}\tau_{kl}\delta_{ij}\tau_{ij}
-\frac{2}{3}\tau_{mm}\delta_{kl}
+\frac{2}{3}\delta_{kl}\tau_{mm}
\right)
=
$$
$$=
\frac{\left\langle \mathcal{F} \right\rangle}{\eta \sigma_{eq}}\frac{3}{2}\frac{1}{J}\frac{1}{J}
\left(
\tau_{kl}-\frac{1}{3}\tau_{mm}\delta_{kl}
\right)=
\frac{\left\langle \mathcal{F} \right\rangle}{\eta \sigma_{eq}}\frac{3}{2}\frac{1}{J}s_{kl}.
$$
Which, in direct notation is equivalent to:
$$
\hat{\mathbf{L}}^p=
\frac{1}{J}\frac{3}{2} \frac{\left\langle \mathcal{F} \right\rangle}{\eta \sigma_{eq}}\mathbf{s}=
\frac{1}{J}\frac{1}{\eta}\frac{3}{2} \frac{\left\langle \sigma_{eq} - \sigma_y \right\rangle}{\sigma_{eq}}\mathbf{s}.
$$
Then, we create an \textit{ad hoc} modification so that the velocity gradient is analogous to the one appearing in the power law crystal plasticity:
$$
\hat{\mathbf{L}}^p=
\frac{1}{J}\frac{1}{\eta}\frac{3}{2} \left(\frac{1}{J}\frac{\sigma_{eq}}{\sigma_y}\right)^n \mathbf{s}.
$$

\bibliographystyle{apsrev4-1}
\bibliography{myreferX}	
\end{document}

%% file: Introduction/intro.tex
\section{Introduction}
\label{sec:intro}

High entropy alloys (HEAs) are multicomponent versions of 
concentrated solid solution alloys (CSAs) that  attracted 
increasing attention throughout recent years
\cite{OSETSKY201865,ShitengZhao}. 
The main reason behind the interest in these materials is their
exceptional properties such as high strength (also at high 
temperature), high hardness and resistance to wear, corrosion
and irradiation, cf. \cite{Kaushik2021,OTTO20132628} and 
references therein. 
Due to these properties, they are considered as candidates for
various applications where the materials have to withstand harsh
conditions, such as future nuclear power plants (generation IV and
fusion) \cite{KURPASKA2022110639}, 
chemical plants as well as aerospace applications and
biomedical implants \cite{ShitengZhao}.
In this paper, we present a continuum plasticity model
considering for the first time the behaviour of a equiatomic
multi-element face centered cubic (FCC) CSA. 
The inherent disorder associated with the lattice distortions
is characterized by an $\alpha$-parameter that quantifies the 
relative importance of an isotropic plastic contribution 
to the model implemented in the finite element method. 
This results in multiple plasticity mechanisms that go beyond 
crystallographic symmetry-based ones.
Besides, molecular dynamics (MD) simulations are performed for 
equiatomic CSA: NiFe, NiFeCr, NiFeCrCo, and Cantor alloys to 
validate the proposed continuum model applied to 
nanoindentation tests obtaining the representative volume
element model by tracking the combined model yield surface. 

The crystal plasticity (CP) theory is a tool that enables us to 
get insight into the active plastic deformation mechanisms (slip,
twinning, transformation) in a continuum setting, leading to a given
mechanical response and microstructure evolution. 
For the material with known properties, it also enables to predict
the mechanical response and microstructure evolution provided that
the initial crystallographic orientation is specified. 
Although the CSAs have been the subject of increasing interest,
some authors \cite{Fang2021} reported recently that crystal 
plasticity (CP) theory has not been used to study equiatomic
CSAs. 
While this is no longer true, simulating the behavior and 
microstructure evolution of CSAs by means of crystal plasticity
theory is still in its nascent stage. In particular, although 
there is a vast number of papers concerning modelling the micro- 
and nanoindentation \emph{of conventional metals and alloys} by means
of crystal plasticity finite element method (CPFEM), cf. e. g. 
\cite{Renner2016,Yao2017,Wang2018,Frydrych2019cmms}, it seems there
is no such study devoted to the CPFEM simulation of the indentation
in chemically complex CSAs like the Cantor Alloy. 
We shall now briefly summarize some CPFEM simulations of CSAs \emph{in
other contexts}. 

Lu et al. \cite{Lu2021jac} applied the classical power law CP
coupled with simple damage model to study the effect of strain
rates and stress states on mechanical behavior and texture 
evolution in polycrystalline CoCrFeNi alloy. 
Gao et al. \cite{Gao2020} applied a power law plasticity with
hardening based on dislocation density evolution to study the
mechanical response and texture evolution of NiCoCrFe deforming
by slip and twinning. 
It should be noted that the study is not strictly a CPFEM where
there is one or more finite elements in each grain. 
Rather, the Taylor assumption was applied in order to merge the
contributions of a number of grains in each integration point. 
A power law crystal plasticity with both isotropic and kinematic 
hardening was applied in \cite{Han2021ijf} in order to simulate 
Fe$_{44}$Mn$_{36}$Co$_{10}$Cr$_{10}$ subjected to 10 cycles of 
tensile-tensile sine load. Maps of cumulative plastic strain 
and dominant slip system were presented. 
The insterstitial HEA Fe$_{49.5}$Mn$_{30}$Co$_{10}$Cr$_{10}$C$_{0.5}$
was studied in \cite{Lu2020jmps, Zhang2022ijp}. 
The modified Orowan equation was used to determine the shear 
rate on each slip system and the evolution of dislocation density
was treated with the modified Kocks-Mecking model. 
The impact of deformation mechanisms on the ratchetting 
strain evolution was analysed in \cite{Lu2020jmps}. 
The influence of temperature and grain size on deformation behavior
was studied in \cite{Zhang2022ijp}.
The tensile behaviour of the polycrystalline Cantor alloy was
studied using fast-Fourier transform (FFT) based crystal 
plasticity model in \cite{Gordon2021}.

Instead of using the finite element method (FEM) or FFT, one can
combine the responses of individual grains by applying 
self-consistent modelling. 
In such a model, each grain is treated as an ellipsoidal
inclusion submerged in a
homogeneous equivalent medium representing the polycrystal.
Tazuddin, Biswas and Gurao \cite{Tazuddin2016} conducted the 
Visco-Plastic 
Self-Consistent (VPSC) \cite{Lebensohn93} model simulation of 
rolling in HEA: MnFeCoNiCu. 
They have concluded that qualitative agreement with the 
experimental texture can be 
obtained only if the plastic deformation is carried out with a 
considerable portion of
the partial $\left\lbrace111\right\rbrace 
\left\langle11\bar{2}\right\rangle$ slip 
alongside the conventional octahedral $\left\lbrace111\right\rbrace 
\left\langle10\bar{1}\right\rangle$ slip.
The VPSC code was used also in \cite{Kaushik2021} in order to
investigate the contributions of twin variants in texture 
development of equiatomic CrMnFeCoNi (Cantor alloy)
subjected to rolling.
Fang et al. \cite{Fang2021} applied a CP model to study the 
behavior of irradiated FeNiMnCr. 
The irradiation was taken into account by introducing the term
stemming from the presence of dislocation loops in the critical
resolved shear stress (CRSS). 
The solid solution strengthening occurring specifically in HEAs
was explained to result from the lattice distortion caused by
differences in shear modulus and atomic size of the atoms
present in the alloy composition. 

On the other hand, strengthening and plastic deformation of HEA are known to be atomic-level phenomena that are usually modeled by using molecular dynamics (MD) simulation to describe the dislocation nucleation and evolution during mechanical testing. The information gained in such MD simulations should be then applied in higher-scale models such as discrete dislocation dynamics or crystal plasticity, which leads us to the multiscale modelling concept. 
Experimental analyses of atomic-level materials phenomena
are highly demanding and need to be validated by 
large-scale atomistic simulation techniques \cite{Choietal,KURPASKA2022110639}.
MD simulations have been widely performed to understand the 
plastic deformation mechanisms of HEA from nanoindentation computational modelling noticing that the length of the dislocation network nucleated
is kept by the geometrically necessary dislocations that
are generated to remove the material from the indenter tip, slowing 
the mobility of the dislocations \cite{ALABDALHAFEZ2019618,QI2021159516}.
A particular advantage of simulating nanoindentation as compared to simulations of other set-ups is the ease at which such simulations can be compared against experimental data. Other experimental techniques at the scale available for MD are typically either impossible or require more effort and are more susceptible to error.
By performing simulations of nanoindentation, the computation 
of the hardness of the material can be done by using the Oliver-Pharr method where this value is higher for Cantor HEA than for single element materials \cite{RUESTES2022111218}.
Moreover, an important feature revealed in MD simulations of spherical nanoindentation is the emergence of a considerable amount of isotropic plasticity \cite{RUESTES2022111218} that is usually compared to pure Ni samples.
Typically, the pure single element metals of 100 orientation show four distinct pile-ups reflecting the crystal symmetry \cite{VARILLAS2017431}. The pile-up symmetry related to a given crystallographic orientation appears naturally not only in MD but also in CP simulations.
On the other hand, the pile-up around an indent in the case of 
chemically complex materials appears to be
almost circular which is typical for an isotropic material, 
as observed for polycrystalline Cantor alloys \cite{QiuYuhang}. 
The four-fold crystallographic symmetry is however still present 
too. 
The limit case of circular pile-up observed in nanoindentation 
can be easily obtained using the isotropic plasticity model, 
e. g. J2 plasticity. 
However, obtaining the response showing traces of both crystal
and isotropic plasticity is to the best of our knowledge not
possible using any continuum mechanics model.

Therefore, the aim of the current contribution is to propose
a continuum model able to reproduce the behaviour of HEAs 
showing both isotropic and crystal plasticity signatures.
The paper is organised as follows. After this introduction, 
in Sec. \ref{sec:MD} we briefly describe \textit{MD simulations} 
performed at three different crystal orientations for a pure Ni
sample and a Cantor alloy. 
The presented results clearly show that the chemical disorder
inherent to this alloy leads a decrease in the pile-up height
variability around an indent.
Next, in Sec. \ref{sec:CP} the \textit{Continuum model of HEAs} 
is  described by the classical crystal viscoplastic model, 
the proposed isotropic viscoplastic model and shows how they
were combined in order to address the effect observed in MD.
It also describes
the details of the finite element simulations performed. 
In Sec. \ref{sec:results} \textit{Results} are presented by
the surface topographies obtained using the continuum model 
compared against the MD results for each orientation. 
Followed by the \textit{Discussion} section (Sec. \ref{sec:disc}) 
highlights the
issues related to the chemical disorder and its relation to 
amorphization, the effect of elastic anisotropy and discusses
the predictions of mechanical response as observed in other
deformation schemes. 
In particular, the yield surface points obtained using crystal
plasticity, isotropic plasticity and the combined 
isotropic-crystal plasticity model are presented.
Finally, \textit{Conclusions} are recapitulated in Sec. 
\ref{sec:concl} where the main 
findings of the paper suggest further research directions
in terms of validation, application and possible extension of
the proposed continuum model. 
It also points out the way to determine its limitations in
terms of scale and atomic structure of a given material.

%% file: MD/MD_method.tex
\section{Molecular Dynamics simulations}
\label{sec:MD}

An atomistic computational model is applied to emulate
nanoindentation test of pure crystalline nickel and equiatomic 
CSAs: NiFe, NiFeCr, NiFeCrCo, and Cantor 
NiFeCrCoMn alloys by the Large-scale Atomic/Molecular
Massively Parallel Simulator (LAMMPS) \cite{THOMPSON2022108171}
and interatomic potentials reported by 
Choi et al. \cite{Choietal}; which 
are based on the second Nearest Neighbor Modified Embedded Atom
Method (2NN-MEAM).
MD simulations first start by defining
the initial energy optimized Ni sample and equilibrated for
100 ps with a Langevin thermostat at 300 K and a 
time constant of 100 fs \cite{Javier2021}. 
This is done until the system reaches a homogeneous 
sample temperature and pressure profile, at a 
density of 8.78 $g/cm^3$. 
A final step is performed by relaxing the prepared sample
for 10 ps to dissipate artificial heat. 

For the crystalline equiatomic CSAs samples, Ni atoms are 
substituted at the required percentage for the CSAs randomly 
in the original FCC Ni sample following a thermalization at
T= 300 K for 2 ns. 
The system is further equilibrated through Monte Carlo (MC) 
relaxation, using a canonical ensemble
with a compositional constraint by performing atomic
swaps, as implemented in LAMMPS.
Then the slab is exposed to the surface boundary 
conditions (2D periodicity) at 300 K into the microcanonical 
emsemble where the number of particles, N, the sample volume,
V, and the system energy, E, are assumed to be constant 
during the NVE molecular dynamics simulation performed for
20 ps (with 1 fs steps)
\cite{KURPASKA2022110639}. 
The dimensions of the simulation box by considering different 
crystal orientations for the 5-element CSA are mentioned in
Table~\ref{tab:MDdata}.

\begin{table}[!b]
\caption{Simulation boxes sizes and number of atoms}
    \begin{tabular}{lrrrrrrrr}
       \hline
        \textbf{Orientation} & \textbf{Size} (\textbf{d$_x$}, 
        \textbf{d$_y$, \textbf{d$_z$}}) [nm] & \textbf{Atoms}  \\
        \hline
        $[001]$ &  33.20 $\hat{x}$  $\times$ 36.09 $\hat{y}$ 
        $\times$ 36.12 $\hat{z}$ & 3 164 800 \\
        $[110]$ &  33.06 $\hat{x}$ $\times$ 36.17 $\hat{y}$   
        $\times$ 38.71 $\hat{z}$ & 3 442 032 \\
        $[111]$ &  33.07 $\hat{x}$ $\times$ 36.12 $\hat{y}$ 
        $\times$ 40.61 $\hat{z}$ & 3 645 720 \\
        \hline
    \end{tabular}
    \label{tab:MDdata}
\end{table}

\begin{figure}[b!]
\centering
\includegraphics[width=0.375\linewidth]{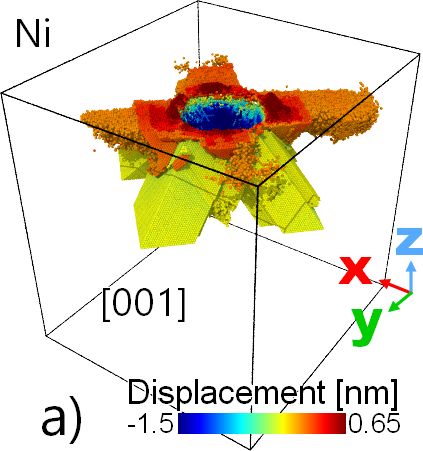} \quad \quad
\includegraphics[width=0.375\linewidth]{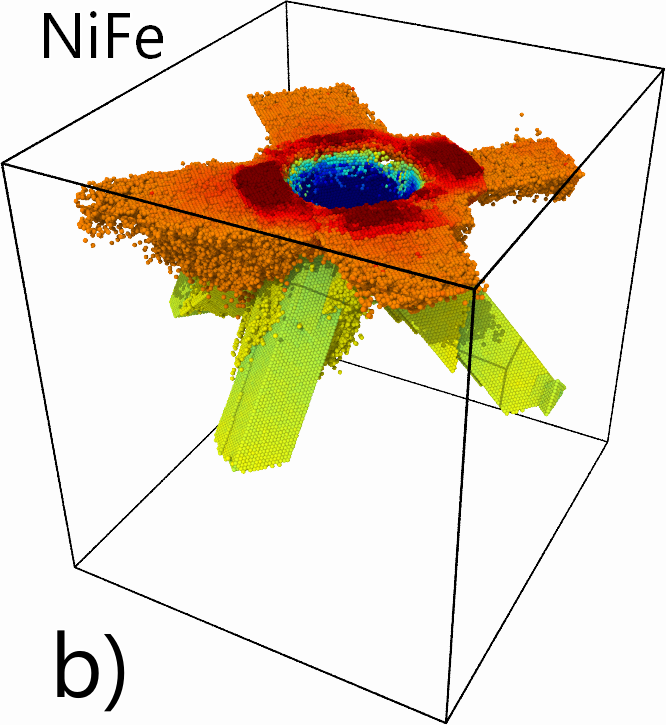} \quad
\includegraphics[width=0.375\linewidth]{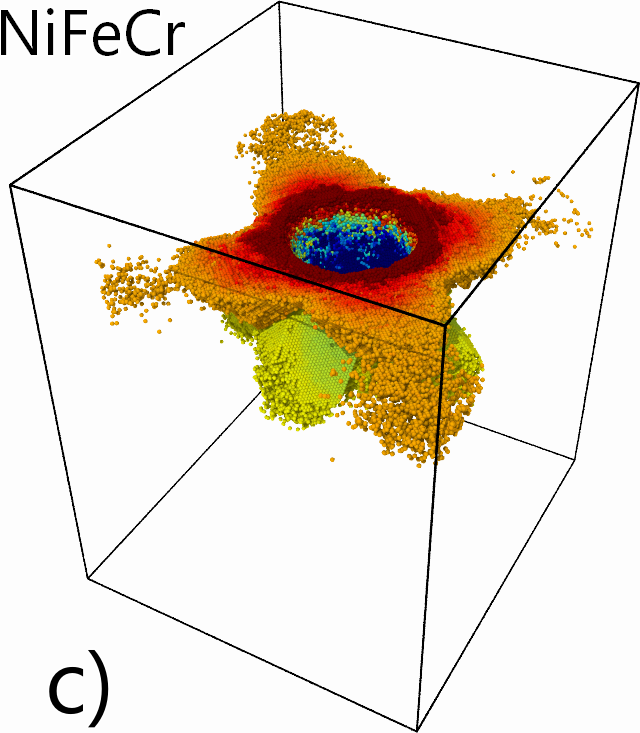}
\quad
\includegraphics[width=0.375\linewidth]{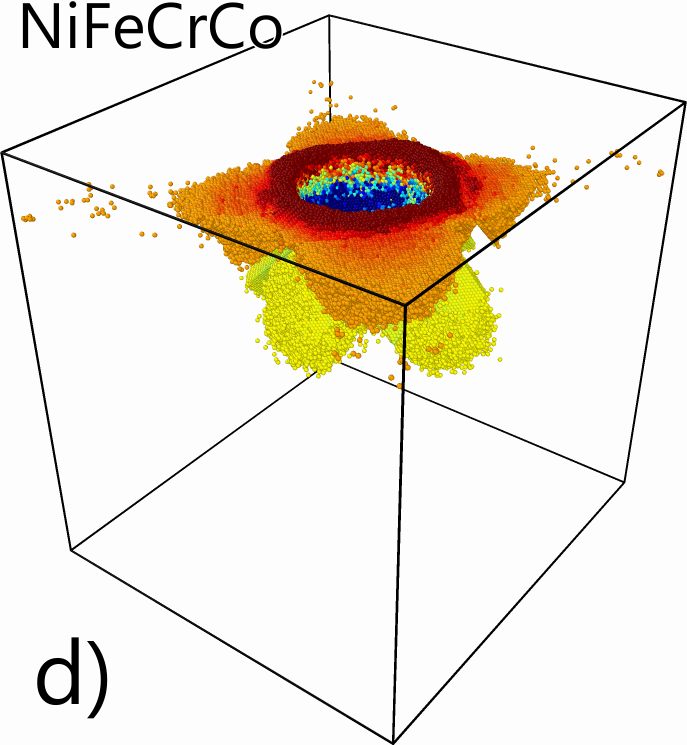}
\quad
\includegraphics[width=0.375\linewidth]{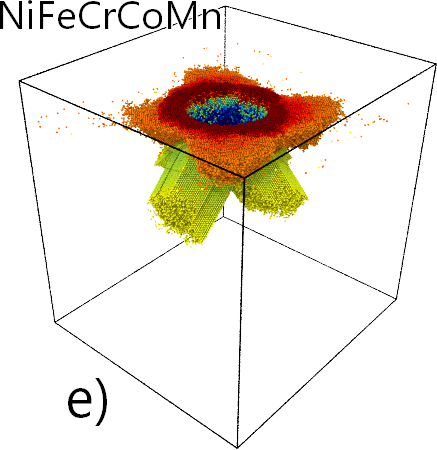}
\caption{Atomic displacements of crystalline Ni in a) and equiatomic
CSA: NiFe in b), NiFeCr in c), NiFeCrCo in d) and Cantor Alloy
(NiFeCrCoMn) in e) on the [001] crystal orientation obtained by MD 
simulations; at the maximum indentation depth of 5nm. 
Slip planes are shown by both samples due to their FCC structure.
However, slip traces on the surface changed due to the chemical
complexity of the Cantor Alloy.}
\label{fig:fig1}
\end{figure}

\subsection{Nanoindentation test}
\label{subsec:nanoindet}

In order to perform MD simulations of nanoindentation, 
the prepared sample is divided into three sections on 
the $z$ direction for setting up boundary conditions. 
In addition, we consider 7 nm vacuum section on the top, 
above the material sample and also the two lowest bottom
layers are kept frozen ($\sim$ 0.02$\times$\textbf{d$_z$}) to 
assure stability of the Mo atoms when
nanoindentation is performed.
A thermostatic region above the already defined frozen one is 
considered to dissipate the generated heat during 
nanoindentation, with a thickness of $\sim$
0.08$\times$\textbf{d$_z$}. 
The rest of the layers the atoms are free to interact with
the indenter tip that modifies the surface structure. 
Periodic boundary conditions are set on the $x$ and $y$ axes
to simulate an infinite surface, while the $z$ orientation
contains a fixed bottom boundary and a free top boundary 
in all MD simulations.

The indenter tip is considered as a non-atomic
repulsive imaginary (RI) 
rigid sphere with a force potential defined as: 
$F(t) = K \left(\vec r(t) - R \right)^2$ where $K = 236$ 
eV/\AA$^3$ (37.8 GPa) is the force constant, and $\vec r(t)$ is 
the position of the center of the tip as a function of time,
with radius $R$. Here, $\vec r(t) = x_0 \hat x + y_0 \hat y + 
(z_0 \pm vt)\hat z$ 
with $x_0$ and $y_0$ as the center of the surface sample on the 
xy plane, the $z_0 = 0.5$ nm is the initial gap between the surface 
and the intender tip moves with a speed $v$ = 20 m/s 
performed for 125 ps with a time step of 
$\Delta t = 1$ fs. 
The maximum indentation depth is chosen to 5.0 nm 
to avoid the influence of boundary layers in the 
dynamical atoms region \cite{KURPASKA2022110639}. 
In Fig. \ref{fig:fig1}, we report results for the atomic
displacement 
of the pure Ni in a) and for equiatomic CSA: NiFe in b) 
\cite{KURPASKA2022110639}, 
NiFeCr in c), NiFeCrCo in d), and Cantor alloy in e) 
on the [001] crystal orientation, 
at the maximum indentation
depth; noticing that single element materials presents slip traces 
with the four pile-ups. 
While the atoms are accumulated around the indenter tip for the 
CSA sample, which is an effect of the lattice mismatch due to the 
chemical complexity of the Cantor alloy. 
Nevertheless, the CSA preserves the FCC geometry as observed for the 
slip planes in Fig. \ref{fig:fig1} for different cases.

Motivated by these results, we decided to model the surface 
morphology of indented CSA by a crystal plasticity model, as 
discussed in the following section.

%% file: Continuum/continuum.tex
\section{Continuum model accounting for chemical disorder}
\label{sec:CP}
In this section we first review the classical crystal viscoplasticity approach. Then we propose the power law isotropic viscoplasticity model. Finally, we suggest the way to combine the abovementioned models in order to account for the behaviour of CSAs which due their inherent lattice-level disorder show the features of both crystalline and isotropic materials.

\subsection{The crystal plasticity model}
The rate-dependent crystal plasticity model with phenomenological hardening 
law implemented in the total Lagrangian setting was already described in a 
number of papers, e. g. \cite{Frydrych2018msmse,Frydrych2019msmse}. 
Here we include its description for the sake of the paper's clarity.

The kinematics follows classical contributions, cf. \cite{Hill72,Asaro77,Asaro85}.
Multiplicative decomposition of the deformation gradient $\mathbf{F}$ into
elastic $\mathbf{F}_e$ and plastic $\mathbf{F}_p$ parts is performed:
\begin{equation}  \label{Multiplicative}
\mathbf{F}=\mathbf{F}_e\mathbf{F}_p,
\end{equation}
and the evolution of the plastic part $\mathbf{F}_p$ is dictated by the 
plastic velocity gradient $\hat{\mathbf{L}}_p$ :
\begin{equation}\label{Fpevolution}
\dot{\mathbf{F}}_p=\hat{\mathbf{L}}_p\mathbf{F}_p.
\end{equation}
The plastic velocity gradient is calculated as a sum of shear rates
$\dot{\gamma}^r$ projected on the $r$-th slip system direction $\mathbf{m}^r_0$ 
and plane normal $\mathbf{n}^r_0$ in the reference configuration:
\begin{equation}\label{Lcrystal}
\hat{\mathbf{L}}^p_{crystal}=\sum_{r=1}^{M}\dot{\gamma}^r\mathbf{m}^r_0\otimes\mathbf{n}^r_0
\end{equation}

The shear rates are obtained using the power law \cite{Asaro85} with the 
reference slip velocity  $v_0$ treated as a model parameter:
\begin{equation}
\dot{\gamma}^r=v_0\mathrm{sign}(\tau^r)\left|\frac{\tau^r}{\tau_c^r}\right|^{n}.
\end{equation}
Here, $n$ is a rate-sensitivity parameter. The Mandel stress tensor
$\mathbf{M}_e$ is projected on a given slip system $r$ in order to calculate
the resolved shear stress (RSS):
\begin{equation}
\tau^r=\mathbf{m}^r_0\cdot\mathbf{M}_e\cdot\mathbf{n}^r_0,
\end{equation}
while the hyper-elastic law is applied to obtain the Mandel stress:
\begin{equation}\label{Mandel}
\mathbf{M}_e=2\mathbf{C}_e\frac{\partial \Psi}{\partial \mathbf{C}_e}.
\end{equation}
Here, the right elastic Cauchy-Green tensor $\mathbf{C}_e=\mathbf{F}^T_e\mathbf{F}_e$ 
is used and the free energy density $\Psi$ is defined in terms of the elastic Lagrangian
strain tensor $\mathbf{E}_e=\frac{1}{2}(\mathbf{C}_e-\mathbf{1})$ and the
four-dimensional stiffness tensor $\mathbb{C}_e$:
\begin{equation}\label{Psi}
\Psi=\frac{1}{2}\mathbf{E}_e\cdot\mathbb{C}_e\cdot\mathbf{E}_e
\end{equation}

Concerning the critical resolved shear stress (CRSS) $\tau_c^r$ on a given 
slip system $r$, its evolution is governed by the exponential 
Voce-type hardening law:
\begin{equation}
\dot{{\tau}_c}^r = H(\Gamma_{crystal})\sum_{s=1}^{M}h_{rs} \left| \dot{\gamma}^s \right|.
\end{equation}
with
\begin{equation}
H(\Gamma)=\frac{d\tau(\Gamma_{crystal})}{d\Gamma_{crystal}},
\end{equation}
and
\begin{equation}
\tau(\Gamma_{crystal})=\tau_0+\left(\tau_1+\theta_1 \Gamma_{crystal} \right)
\left(1-\exp\left({-\Gamma_{crystal} \frac{\theta_0}{\tau_1}}\right) \right)
\end{equation}
Where the accumulated plastic shear is defined as follows:
\begin{equation}\label{Gcrystal}
\Gamma_{crystal}=\int \dot{\Gamma}_{crystal} dt,
\hspace{0.2cm}
\dot{\Gamma}_{crystal}=\sum_{r}{|\dot{\gamma}^r|}
\end{equation}
The latent hardening parameter $h_{rs}$ is equal to 1 for $r=s$ (self-hardening),
whereas for $r \neq s$ (latent hardening), it is equal to $q_0$ on coplanar 
systems and $q$ on non-coplanar systems.
The model parameters for pure Ni were taken from \cite{Lewandowski2018} and are
listed in Tab. \ref{CPparam}.

\begin{table*}[!t]
\centering
\caption{Elastic constants and crystal plasticity model parameters for pure Ni \cite{Lewandowski2018}.}
\label{CPparam}
\begin{tabular}{lrrrrrrrrrr}
\hline
$C_{11}$   & $C_{12}$   & $C_{44}$ & $\tau_0$ & $\tau_1$ & 
$\theta_0$ & $\theta_1$ & $q$      & $q_0$    & $v_0$ & $n$  \\
\hline
GPa &  GPa   & GPa & GPa  & GPa  & GPa
  &  GPa &     &     &  & \\
\hline
246.5 & 147.3 & 124.7 & 0.008 & 0.142 & 0.240 & 0.0075 & 1.4 & 1.0 & 0.0001 & 10\\
\hline
\end{tabular}
\end{table*}

\subsection{The isotropic plasticity model}
In order to simulate the nanoindentation with isotropic distribution of plastic
shear, some J2-based model (as the one in section 2.1 of \cite{Frydrych2019cmms})
could have been applied. However, we wanted to have the model directly analogous
to the CP model so that the combined model can then be built (see the next section).
The structure of the model is thus similar to the CP model presented above. 
The differences are described in the following.

Equations \ref{Multiplicative} and \ref{Fpevolution} still hold. 
We propose the model of the isotropic contribution to be analogous to the
finite strain version of the viscoplastic Perzyna-type model. 
However, we replace the yield surface and its derivative with the power law. 
We thus obtain (details of the derivation are presented in the Appendix):
\begin{equation}\label{Lamorph}
\hat{\mathbf{L}}^p_{isotr}=\frac{1}{\eta}
\left(\frac{1}{J}\right)^{n+1}
\left(\frac{\sigma_{eq}}{\sigma_{y}} \right)^n
\mathbf{s},
\end{equation}
where $\eta$ is the viscosity parameter, $n$ is the rate sensitivity exponent (the same as in the crystal plasticity model), $\sigma_{eq}$ is the equivalent Huber-Mises stress, $\sigma_{y}$ is the yield stress and $\mathbf{s}$ is the stress deviator. 

Equations \ref{Mandel} and \ref{Psi} still hold. The linear-exponential hardening describes the evolution of the yield stress:
\begin{equation}
\sigma_y = \sigma_{y0} + K_{h} \Gamma_{isotr} + R_{inf} \left(1 - e^{-\delta_h \Gamma_{isotr}}\right),
\end{equation}
where $\sigma_{y0}$, $K_{h}$, $R_{inf}$ and $\delta_h$ are the hardening parameters. The model parameters for pure Ni callibarated in such a way the the mechanical response is equivalent to the crystal response are presented in Tab. \ref{AMparam}.

\begin{table}[!b]
\centering
\caption{Elastic and plastic parameters of the isotropic model for pure Ni.}
\label{AMparam}
\begin{tabular}{lrrrrrrr}
\hline
$E$ &$\nu$ &$\sigma_{y0}$ &$K_h$ &$R_{inf}$ &$\delta_h$ &$\eta_{rel}$ &$n$\\
\hline
GPa &   -   & GPa          & GPa   & GPa      & -  &$(\rm{GPa}\cdot s)^{-1}$ & -\\
\hline
220 & 0.31 & 0.012 & 0.1 & 0.04 & 22 & 1000 & 10\\
\hline
\end{tabular}
\end{table}

$\Gamma_{isotr}$ is the accumulated plastic shear which is defined as follows:
\begin{equation}\label{Gamorph}
\Gamma_{isotr} = \sqrt{\frac{2}{3} \mathbf{E}_p . \mathbf{E}_p},
\end{equation}
where $\mathbf{E}_p = \frac{1}{2}\left(\mathbf{F}_p^T\mathbf{F}_p-\mathbf{I}\right)$ is the plastic Lagrangian strain tensor.

\subsection{The combined isotropic-crystal plasticity model}\label{combined}
Finally, the crystal and isotropic plasticity models are combined in order to account for the observed behaviour of HEAs. The degree of isotropy is described by the parameter $\alpha$. Therefore, the velocity gradient is now defined as follows:
\begin{equation}
\hat{\mathbf{L}}^p_{HEA} = \alpha \hat{\mathbf{L}}^p_{isotr} + (1-\alpha) \hat{\mathbf{L}}^p_{crystal},
\end{equation}
where $\hat{\mathbf{L}}^p_{isotr}$ and $\hat{\mathbf{L}}^p_{crystal}$ have been defined in equations \ref{Lamorph} and \ref{Lcrystal}, respectively. Also the hyper-elastic law \ref{Mandel} is still valid but the elastic stifness tensor is now computed as:
\begin{equation}\label{CHEA}
\mathbb{C}_{HEA} = \alpha \mathbb{C}_{isotr} + (1-\alpha) \mathbb{C}_{crystal}.
\end{equation}
The hardening contributions are calculated separately for the isotropic and crystal parts.

\subsection{The finite element simulation details}
The material models described above have been implemented using the AceGen software \cite{Korelc2002,Korelc08}, which takes advantage of the automatic code generation, automatic differentiation and automatic expression optimization techniques. In the case of the crystal plasticity and the combined model, 12 $\left\lbrace111\right\rbrace\left<110\right>$ FCC slip systems were used. The finite element simulations have been carried out using the AceFEM software \cite{Korelc2002,Korelc08} applying 8-noded hexahedral elements. The F-bar method \cite{SouzaNeto96} is applied for the sake of the robustness of the implementation when the nearly incompressible behaviour of the material is enforced in the finite deformation regime.

\begin{figure}[!t]
\includegraphics[width=.44\linewidth]{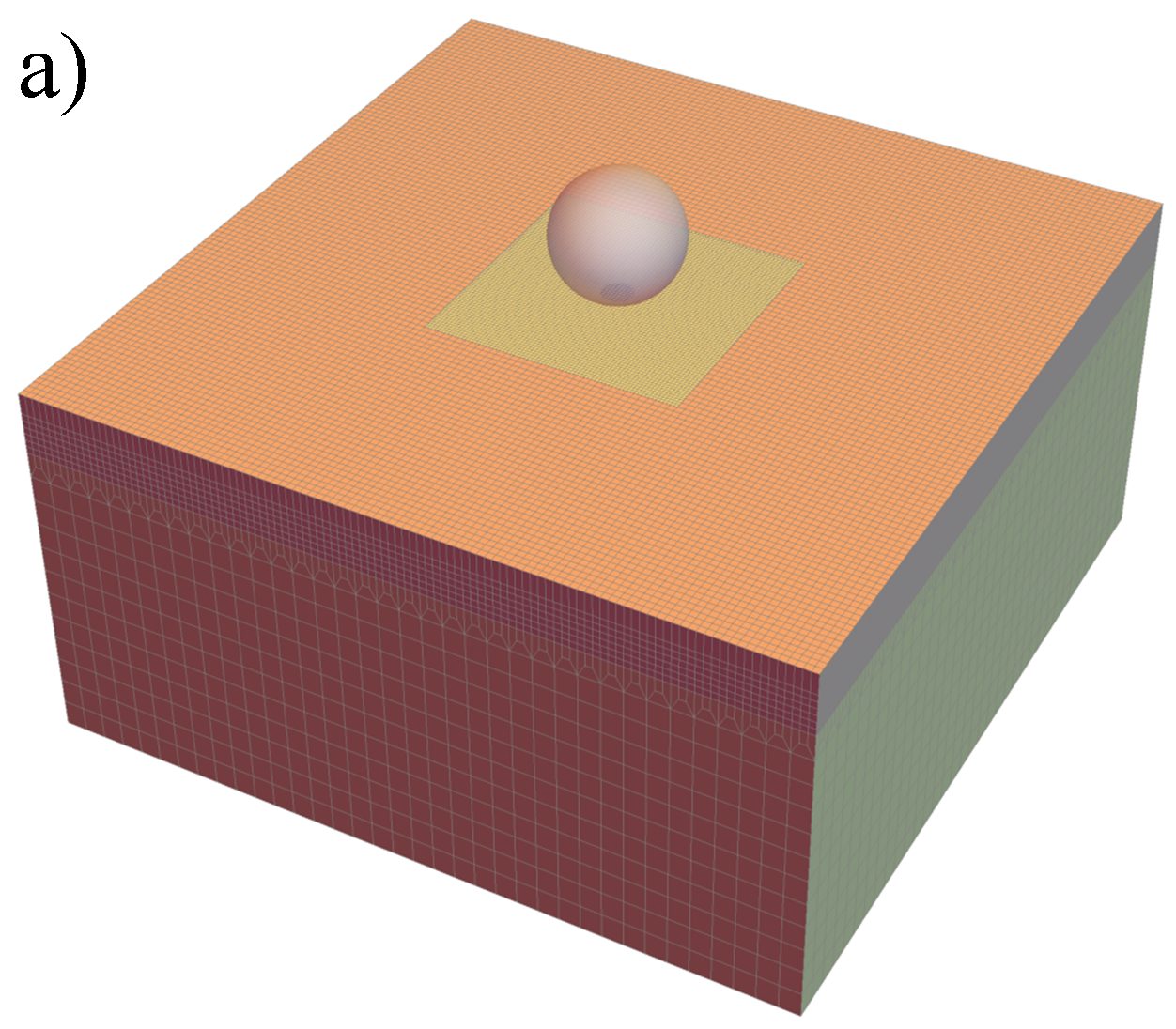}
\includegraphics[width=.54\linewidth]{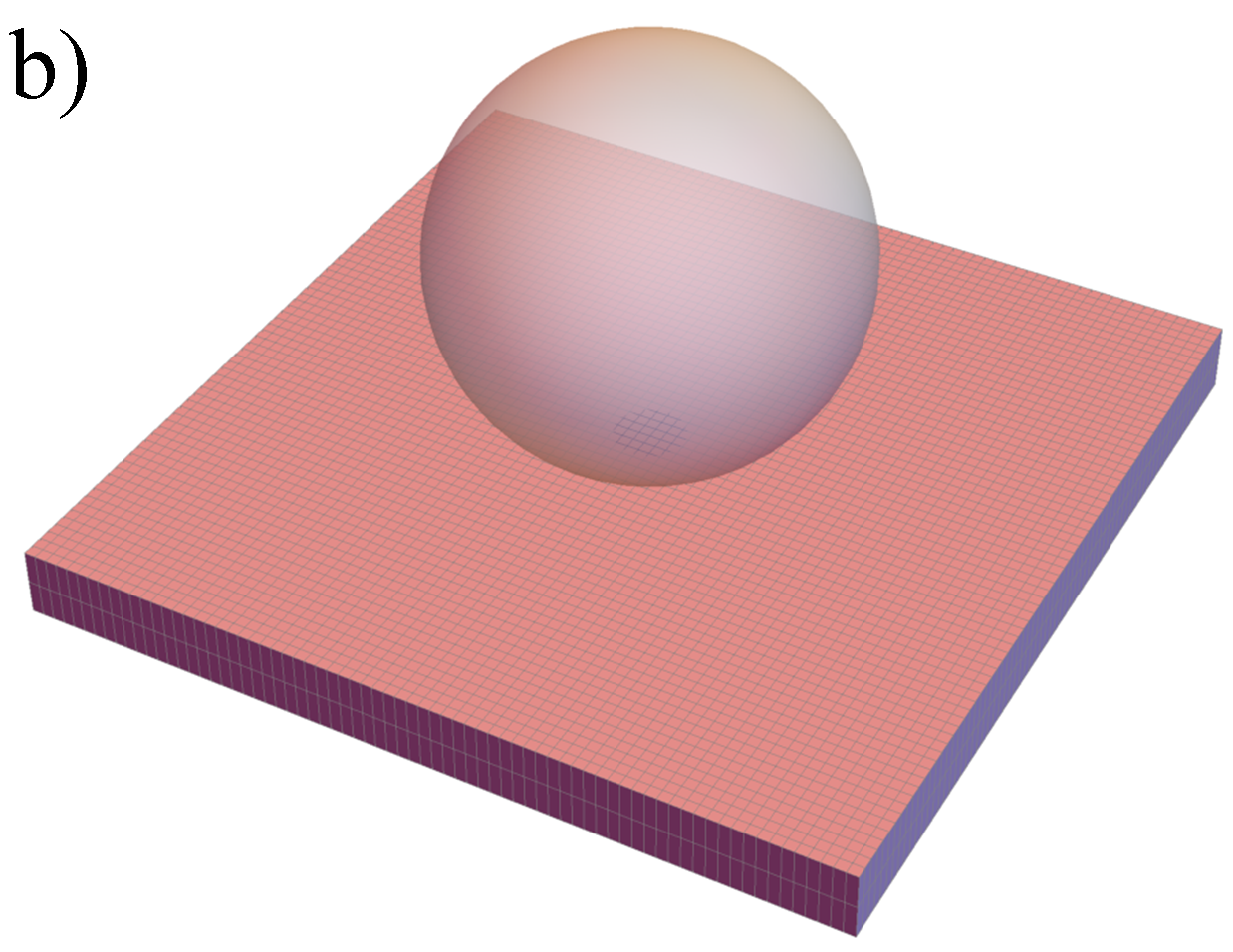}
\caption{The finite element mesh used in all the FEM simulations reported here together with the visualization of the indenter: a) the whole mesh, b) the most refined region.}
\label{mesh}
\end{figure}

The mesh used to perform the simulations is similar to the one used in \cite{Frydrych2019cmms} and shown in Fig. \ref{mesh}. The applied mesh density is not necessary from the point of view of the solution convergence (see \cite{Frydrych2019cmms} for mesh convergence analysis) but it is done in order to get an excellent resolution of the surface topography maps. The dimensions of the whole mesh along x and y directions are 60 nm and the height is equal to 30 nm. In total, the mesh is built using 121,824 linear hexahedral elements.
The most refined region shown in Fig. \ref{mesh}b has x and y dimensions equal to 20 nm and is 1.67 nm thick. It was divided into 72 x 72 x 2 finite elements. The refined region is connected with the rest of the mesh using a hanging nodes formulation described in section 2.4 of \cite{Frydrych2019cmms}. 

The essential boundary conditions are applied as follows. The deformation of the lateral surfaces in the normal direction is set to zero. The bottom surface is forced to move in the positive z direction up to $h_{max}=5nm$ and the position of the indenter is fixed. This is of course equivalent to fixing the bottom surface along the z direction and moving the indenter in the negative z direction but the former approach was easier to implement.

The nanoindentation was simulated using the contact surface element implemented in AceGen. The spherical indenter is treated as a rigid ball. The 4-noded quadrilateral element with Lobatto quadrature is applied. 
In order to exactly fulfill the impenetrability condition:
\begin{equation}\label{impen}
g_N \geq 0 \land t_N \leq 0 \land t_N g_N = 0,
\end{equation}
the augmented Lagrangian formulation (cf. \cite{LengiewiczPhD,Stupkiewicz2010,Frydrych2019cmms}) was used. In Eq. \ref{impen} $g_N$ denotes the normal gap and $t_N$ is the contact pressure. Similarly as in \cite{Frydrych2019cmms}, friction was neglected here.

%% file: Results/results.tex
\section{Results}
\label{sec:results}

\begin{figure*}
\begin{tabular}{rc}
\includegraphics[width=.4\linewidth]{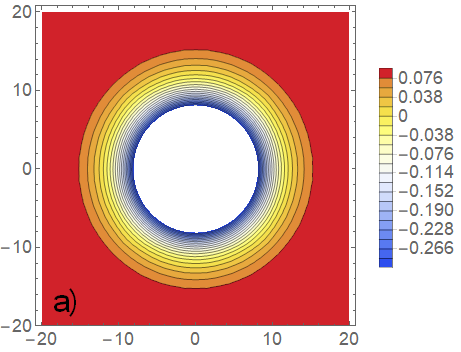}&
\includegraphics[width=.4\linewidth]{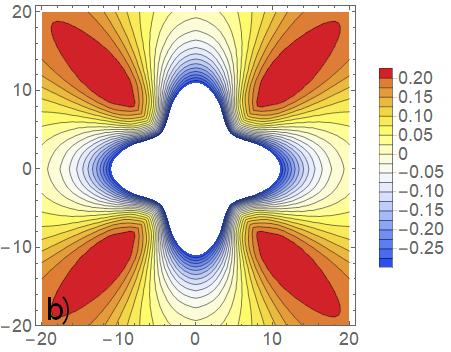}\\
\includegraphics[width=.4\linewidth]{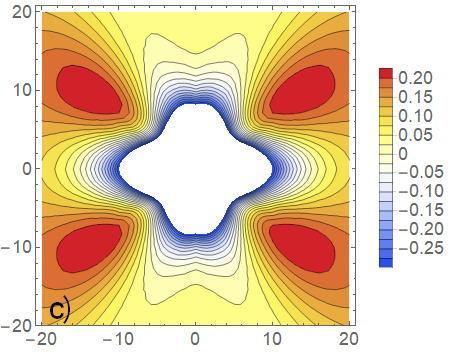}&
\includegraphics[width=.4\linewidth]{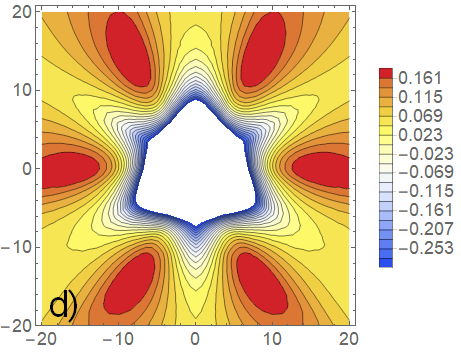}\\
\end{tabular}
\caption{The surface deformation after indentation simulated using FEM with isotropic model (a) and crystal plasticity for [100] (b), [110] (c) 
and [111] (d) orientations. The scale is in nm.}
\label{Fig2}
\end{figure*}

\begin{figure}[b!]
    \centering
    \begin{tabular}{ccc}
    \includegraphics[width=0.35\linewidth]{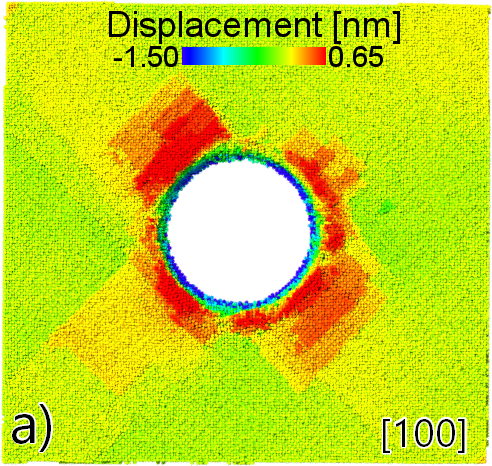}&
    \includegraphics[width=0.345\linewidth]{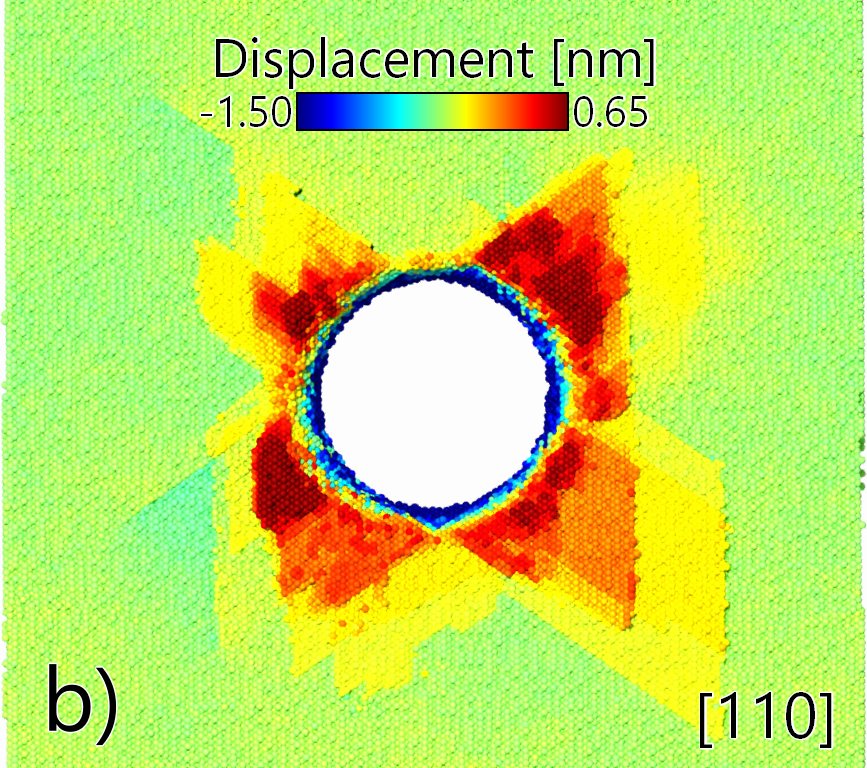}&
    \includegraphics[width=0.265\linewidth]{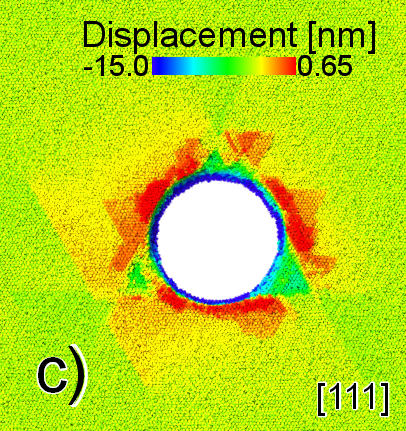}\\
    \end{tabular}
    \caption{Atoms displacements at the maximum indentation depth for a crystalline pure Ni 
    oriented on [100] in a), on [110] in b), and [111] in c), obtained in MD simulation.}
    \label{Fig3}
\end{figure}

Fig. \ref{Fig2} shows the surface normal displacement maps
obtained by using the isotropic plasticity model and the
crystal plasticity model for 100, 110 and 111 orientations 
for a pure Ni sample. 
Since the spherical tip is considered, the surface 
topography resulting from the isotropic model simulation
does not show any preferred direction. 
In addition, the symmetries predicted by the CP model 
directly reflect the crystallographic symmetry for a given 
crystallographic orientation.
The results of the MD simulations with the same tip radius
and maximum indentation depth are shown in Fig. \ref{Fig3}. 
Since different methods were used, an exact quantitative agreement
was not expected. However, the qualitative agreement in 
terms of the type of pile-up symmetry can be observed. 
In particular, one can see the four-fold symmetry for 
orientation 100 (cf. Fig. \ref{Fig2}b and \ref{Fig3}a), 
two perpendicular mirror planes with four pile-ups in the
case of orientation 110 (cf. Fig. \ref{Fig2}c and Fig.
\ref{Fig3}b) as well as the three-fold symmetry with mirror
planes for orientation 111 (cf. Fig. \ref{Fig2}d
and Fig. \ref{Fig3}c).

\begin{figure}
\flushright
\includegraphics[width=.95\linewidth]{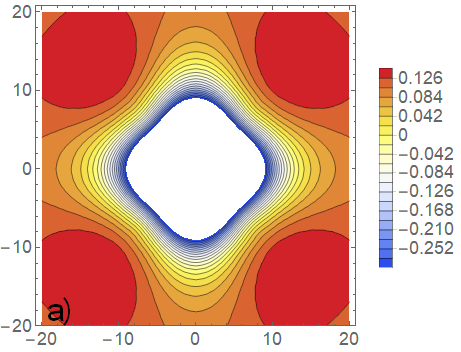}\\
\centering
\includegraphics[width=.65\linewidth]{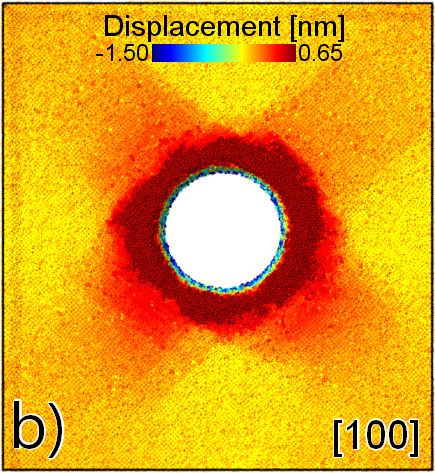}
\caption{The surface deformation after for orientation 100 with a) FEM simulation with the combined HEA model and b) MD simulations. The numbers are in nm.}
\label{Fig4}
\end{figure}

As it was already discussed in the MD simulation section, 
in CSAs, the deformation by glide on crystallographic slip
systems is less pronounced than in classical metals and alloys
(cf. Fig. \ref{fig:fig1}). In particular, the surface topography
in nanoindentation
is affected by crystal symmetry to a lesser extent and presents
some degree of isotropy. 
This can be especially seen when comparing the MD results
of the high entropy alloy with
pure Ni for orientation 100 (\ref{Fig4}b and \ref{Fig3}a), 110 (\ref{Fig6}b and 
\ref{Fig3}b) and 111 (\ref{Fig7}b and \ref{Fig3}c). 
Accounting for this effect was a goal of the combined isotropic-crystal plasticity model
developed in section \ref{combined}. Fig. \ref{Fig4}a shows the surface topography after
indentation predicted using this model for orientation 100 with equal contributions of
isotropic and crystal plasticity ($\alpha=0.5$). Comparing with Fig. \ref{Fig2}a and
\ref{Fig2}b one can clearly see the origin of this topography. This is in qualitative 
agreement with the MD result for HEA shown in \ref{Fig4}b. 

\begin{figure}
\flushleft
\includegraphics[width=.9\linewidth]{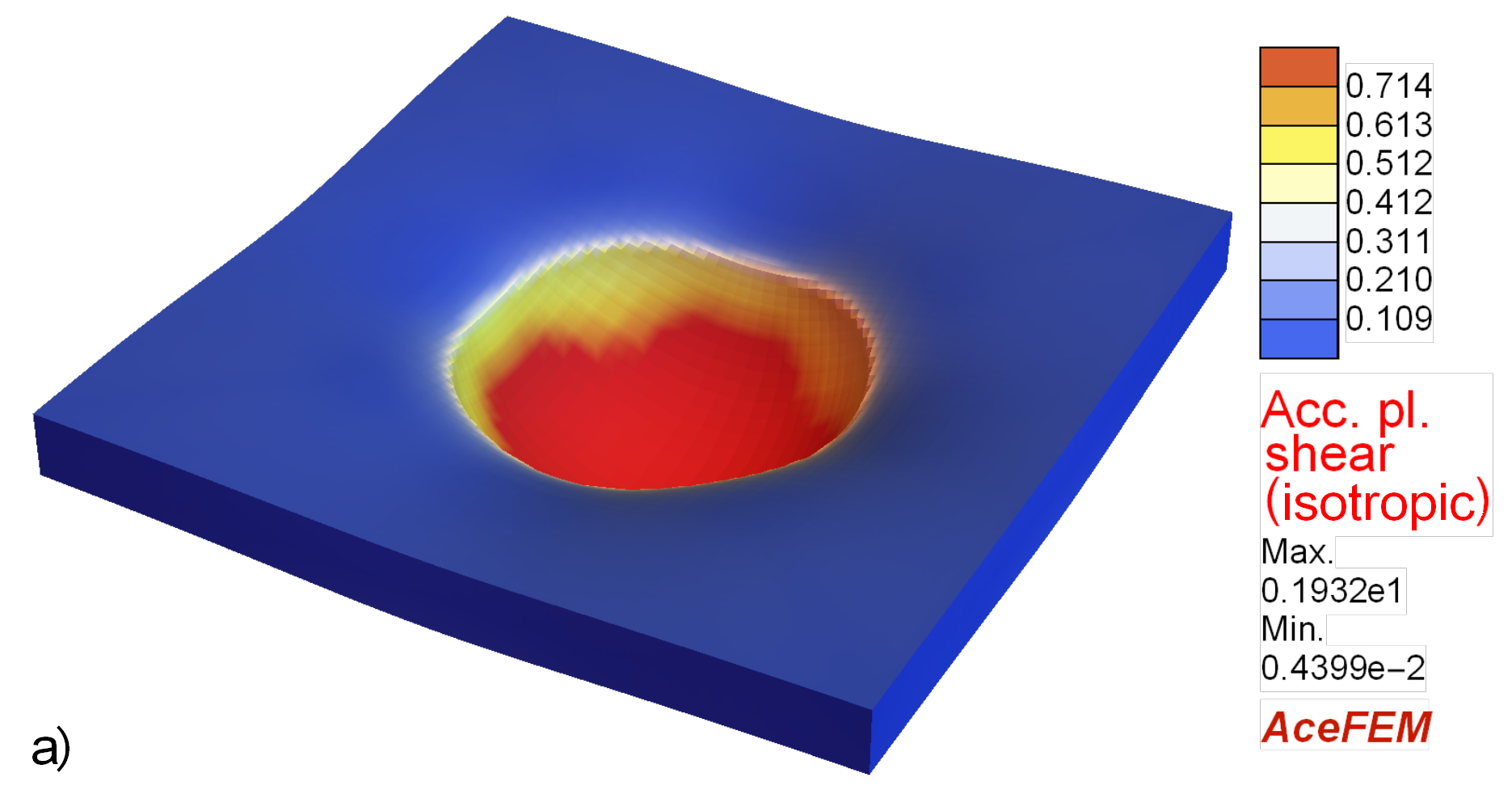}\\
\includegraphics[width=.9\linewidth]{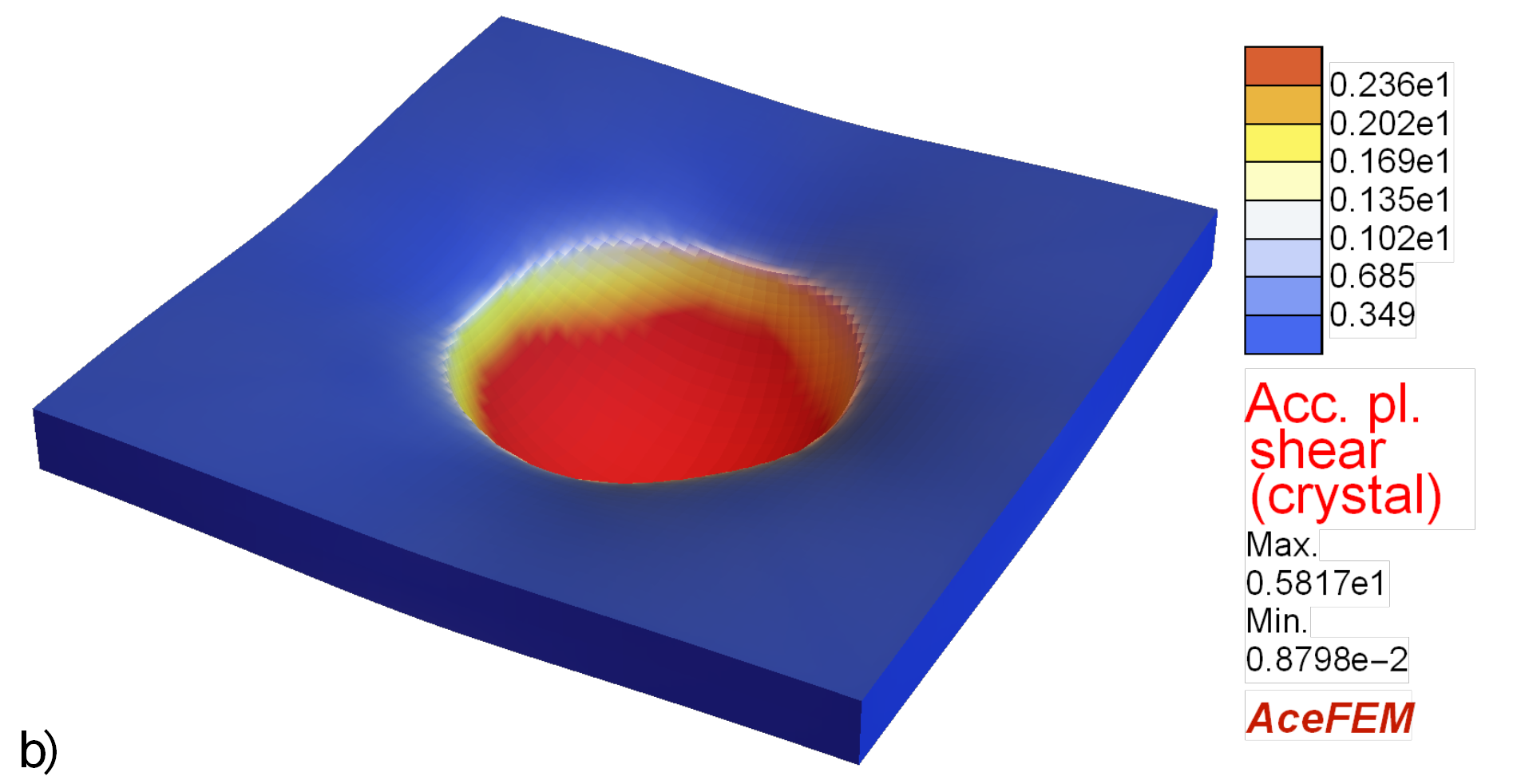}
\caption{The accumulated shear coming from a) isotropic (cf. Eq. \ref{Gamorph}) and b) crystal (cf. Eq. \ref{Gcrystal}) plastic shearing for orientation 100 obtained in FEM simulation with the combined HEA model.}
\label{Fig5}
\end{figure}

\begin{figure}
\flushright
\includegraphics[width=.95\linewidth]{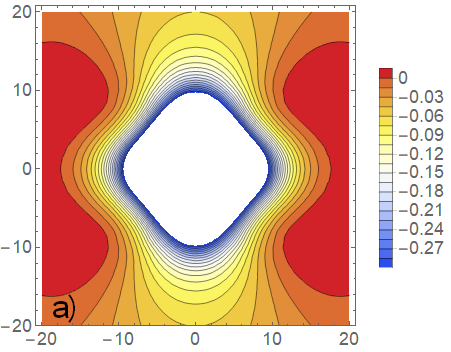}\\
\centering
\includegraphics[width=.70\linewidth]{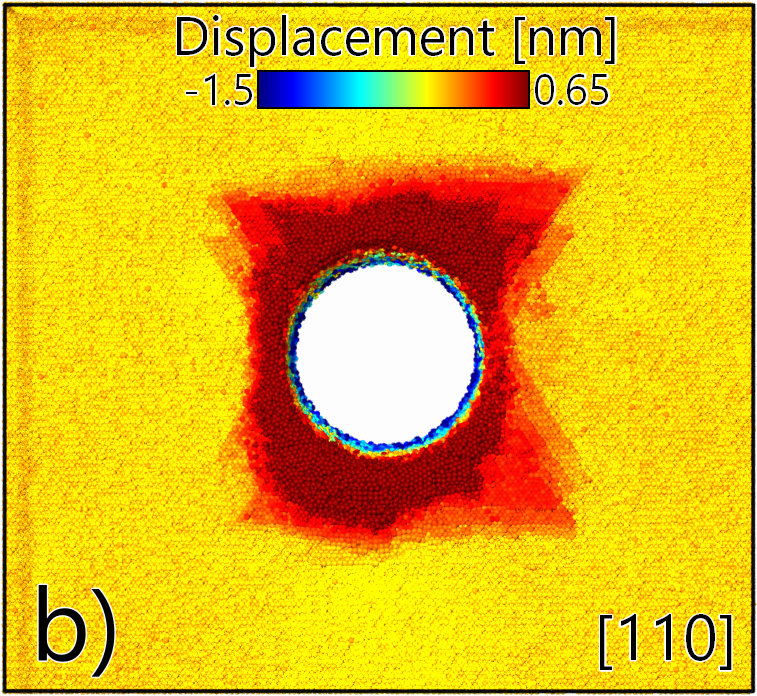}
\caption{The surface deformation after indentation for orientation 110 with a) FEM simulation with the combined HEA model and b) MD simulations. The numbers are in nm.}
\label{Fig6}
\end{figure}

\begin{figure}
\flushright
\includegraphics[width=.95\linewidth]{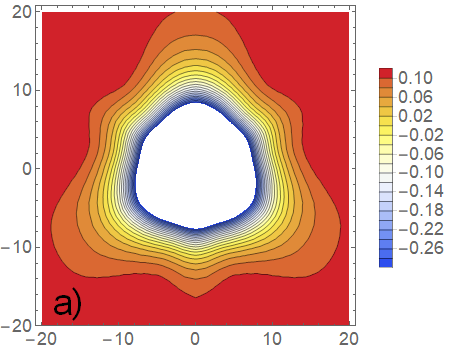}\\
\centering
\includegraphics[width=.65\linewidth]{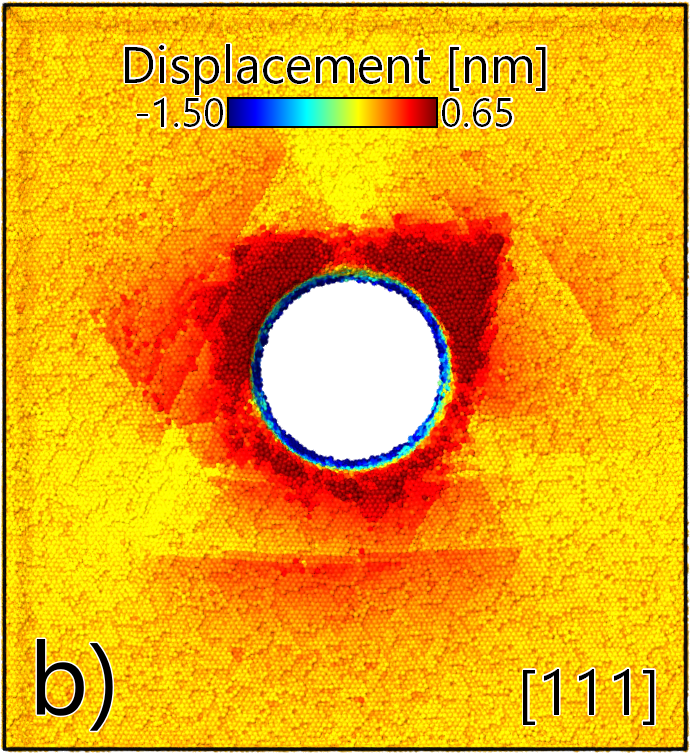}
\caption{The surface deformation after indentation for orientation 111 with a) FEM simulation with the combined HEA model and b) MD simulations. The numbers are in nm.}
\label{Fig7}
\end{figure}

Fig. \ref{Fig5} presents the amount of accumulated plastic shear $\Gamma$ divided into a) isotropic (cf. Eq. \ref{Gamorph})  and b) crystal plasticity (cf. Eq. \ref{Gcrystal}) contributions in the combined model for orientation 100. One can see that the observed effect of isotropization is indeed the result of simulataneous contributions of isotropic and crystal plasticity.

Figures \ref{Fig6} and \ref{Fig7} show the results analogous to \ref{Fig4} for orientations 110 and 111, respectively. These pictures confirm that the observed effects are not related to any particular crystallographic orientation, which was expected.

%% file: Discussion/discussion.tex
\section{Discussion}
\label{sec:disc}
Wang et al. \cite{Wang2021deformation} studied the crystalline-to-amorphous phase transformation ahead of the crack tip in the NiFeCrCoMn Cantor alloy. Based on MD simulations with binary Cu-Al potentials, they concluded that the mechanism of transformation is the dislocation tangles development which finally lead to emergence of the amorphous phase. The severe increase of the dislocation density is due to high lattice resistance which limits the dislocation motion. The authors enlist also other possibilities to obtain amorphous phase by plastic deformation reported in the literature, such as cold rolling or high pressure torsion, where the transformation mechanism is similar.
Moreover, as noted in \cite{ALABDALHAFEZ2019618}, the same amount of plastic strain in the multi-component alloy, as compared to pure metal, leads to a higher dislocation density in a smaller volume. This can finally lead to amorphization. Although in \cite{ALABDALHAFEZ2019618} the amorphization was not observed in the case of the Cantor alloy subjected to indentation, this can be due to a different MD potential than the one used in \cite{Wang2021deformation}.

It is interesting to link these results to our findings. Note that the severe lattice distortion inherent in the structure of HEAs leads to a decreased peak intensity and greater peak width (cf. \cite{ALABDALHAFEZ2019618,Mishra2021}) and high lattice friction leads to increased dislocation density. Even if the combination of those two effects does not lead to a fully amorphous system, it is expected that the system's behaviour will be impacted by this added disorder so using the combined isotropic-crystal plasticity model as proposed here seems to be justified.

\begin{figure}[t!]
   \centering
   \includegraphics[width=0.45\textwidth]{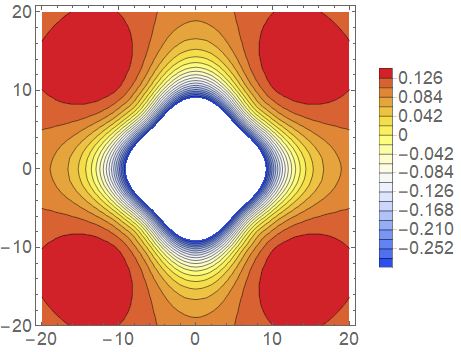}
   \caption{(Color on-line). The surface deformation after FEM simulation of indentation for
   orientation 100 with the combined HEA model and elastic constants: $C_{11}=170$GPa,
   $C_{12}=100$GPa and $C_{44}=143$ GPa \cite{Zhang2018}.}
   \label{FigElAn}
\end{figure}

In our isotropic-crystal plasticity model, we have proposed that both the elastic and
plastic deformation consists of isotropic and crystal components as this appeared as the
most natural way of building the model. However, it was reported that elastic anisotropy
of the Cantor alloy is actually higher than in the case of Nickel. 
In particular, the Zener anisotropy parameter
\begin{equation}
    A_{Zener} = \frac{2C_{44}}{C_{11}-C_{12}}
\end{equation}
for Ni is equal to 2.51 \cite{Lewandowski2018}, whereas for the 
Cantor alloy the reported values are 2.88 \cite{Gordon2021},
4.09 \cite{Zhang2018} and 4.15 \cite{Zaddach2013}. 
The Zener parameter calculated for the elastic stiffness
tensor calculated in Eq. \ref{CHEA} and parameters reported
in Tabs. \ref{CPparam} and \ref{AMparam} is equal to 1.56
which is obviously much lower. Therefore, we decided to 
verify if the elastic anisotropy has any effect on the
nanoindentation results presented here. 
Fig \ref{FigElAn} shows the surface deformation for 
orientation 100 simulated with full elastic anisotropy
taking elastic constants equal to $C_{11}=170$GPa, 
$C_{12}=100$GPa and $C_{44}=143$ GPa \cite{Zhang2018} 
(which yields $A_{Zener}=4.09$).
Comparing with \ref{Fig5}a it can be concluded that
no discernible difference can be seen and thus the 
results presented in the Results section are valid 
even though the high elastic anisotropy of HEAs was 
not taken into account.

Apart from indentation, in the vast majority of papers reporting mechanical tests of HEAs the response was analysed by uniaxial tension \cite{Gao2020,Fang2021,Lu2021jac,Zhang2022ijp}. Other types of tests include cyclic tension-compression \cite{Lu2020jmps}, fatigue \cite{Han2021ijf} and shear-tension experiments \cite{Lu2021jac}.
However, it appears no one has systematically studied the yield surface of any HEA. Therefore, we provide the yield points calculated using our model in order to trigger the discussion and in order to show how the developed model works. They were calculated with a representative volume element (RVE) consisting of 1000 elements with random crystallographic orientations subjected to periodic boundary conditions (see \cite{Frydrych2018msmse,Frydrych2019msmse} for details) with various $\frac{F_{22}}{F_{11}}$ ratios so as to achieve various points in the $\sigma_{1}-\sigma_{2}$ principal stress plane. The simulation was stopped when the accumulated plastic shear $\Gamma$ was equal to 0.002 and the stress tensor values were recorded at that point. 

\begin{figure}[t!]
   a)\includegraphics[width=0.45\textwidth]{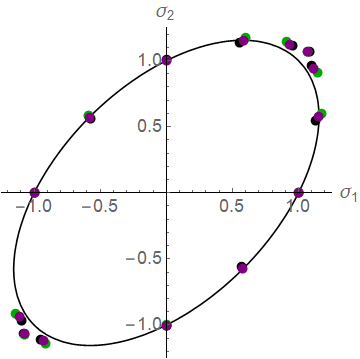}
   b)\includegraphics[width=0.45\textwidth]{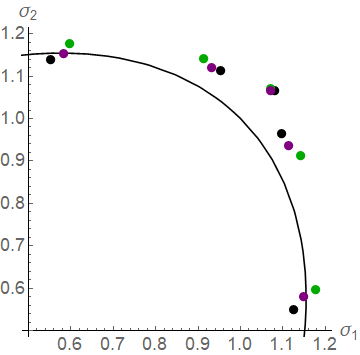}
   \begin{center}
   \includegraphics[width=0.35\textwidth]{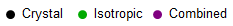}
   \end{center}
   \caption{(Color on-line). Normalized yield points in the $\sigma_1 - \sigma_2$ plane obtained in FEM simulations with all three continuum models together with the normalized J2 yield surface: a) all calculated points, b) close-up of the interesting region. The single parameter $\alpha$ allows to obtain the yield points intermediate with respect to the isotropic and crystal plasticity models.}
   \label{FigYield}
\end{figure}

The normalized yield points (stress values divided by $\sigma_{1}$ at yield at uniaxial tension) are shown in Fig. \ref{FigYield}. The J2 yield surface is also provided. The highest differences appear in the biaxial state, therefore a close-up of this region is provided in Fig. \ref{FigYield}b). One can clearly see that the power-law isotropic plasticity points (in green) are the most ooutward and the crystal plasticity model (black points) are the most inward. It should be however noted that maximum differences do not exceed 3\%. The combined isotropic-crystal plasticity model provides the points that lie in-between the two which was expected. This proves that the model leads to reasonable predictions also for the polycrystalline case. Moreover, it can be concluded that the position of the yield point obtained with the combined model is driven by the value of the single parameter $\alpha$.

Qiu et al. \cite{Qiu2019} reported that compression in the case of Cantor alloy is associated with much higher stress than tension. Interestingly, even the elastic part of the stress-strain curve in compression and tension do not match each other. It should be noted that the geometry of tensile and compressive specimens in \textit{op. cit.} was considerably different so it is hard to say if the effect is real or geometry-related. Moreover, no details on the procedure to avoid barelling in the case of compression specimens was supplied. On the contrary, Jang et al. \cite{Jang2016} who described their procedure to avoid barrelling in detail, did not observe any considerable tension-compression asymmetry while studying the same alloy composition. Therefore, we do not introduce any tension-compression asymmetry in our continuum model, bearing in mind that such asymmetry can be easily introduced in the case any real HEA to be studied presents such a behaviour.


%% file: Conclusions/conclusion.tex
\section{Conclusions}
\label{sec:concl}

In this paper, a continuum model of concentrated solid solutions has been proposed. 
Using the single parameter $\alpha$, the model combines the features stemming from both crystalline symmetries, as well as 
an isotropic disorder-induced contribution, in both elastic and plastic regimes. 
Definition of the model in the finite deformation regime enables us to perform simulations
of nanoindentation tests. 
The predictions of the combined continuum model have been shown to be qualitatively 
consistent with atomistic (MD) simulations both in terms of the resulting surface topography
maps and load-displacement curves. 
To the best of authors' knowledge, this is a first study where lattice disorder, inherent 
in HEAs, is taken into account in the continuum crystal plasticity based model. 
In addition, such issues as amorphization and its relation to atomic disorder, as well as 
the effect of elastic anisotropy were discussed. 
Furthermore, the predictions of the normalized yield surface points obtained by
subjecting a polycrystalline RVE to various periodic boundary conditions were supplied.

For validation, we have shown that the model predictions are qualitatively consistent with
atomistic simulations, but it should be further validated by comparing with other lower
scale models, such as discrete dislocation dynamics, as well as with experimental data. 
In particular, systematic investigations using nanoindentation and atomic force microscopy
of various compositions starting from pure Ni and equiatomic CSAs as in Fig. \ref{fig:fig1} 
would provide adequate results to validate the proposed approach. 
Other ways of validating the reported model could be provided by experimental
investigation of normalized yield surface in CSAs as compared
to pure metals (cf. Fig. \ref{FigYield}).

The model has been validated through molecular dynamics simulations of equiatomic 
CSA: NiFe, NiFeCr, NiFeCrCo, and Cantor alloys, possibly the most
intensively studied HEAs to date. 
In the future, the proposed continuum model should be compared also against other CSAs. 
For this purpose, the proposed continuum model requires the development of an additional
set of parameters that are in general different for each material studied. 
Here, as only the qualitative agreement with MD was sought, the parameter calibration 
step for the combined model was omitted (the parameters established separately for the
crystal and isotropic plasticity models were used also in the continuum model). 
However, seeking for a quantitative agreement with experimental data, it should be
calibrated. We have recently conducted the optimization of crystal plasticity parameters
using evolutionary algorithm in a similar spirit to what was done in \cite{Chakraborty2017},
where the Nelder-Mead algorithm was applied for this task. In the future, we will apply 
our optimization approach in the CSA context so that the model is in quantitative agreement
with the experimental data both in terms of the measured surface topography and the
stress-strain response.

As pointed out in \cite{Antillon2020}, the random occupation of lattice sites by 
constituent atoms (as observed in the Cantor alloy) is not common in other CSAs.
Rather, the atomic configurations that are thermodynamically most stable typically
present some chemical short range order (C-SRO). It is therefore interesting to check
how C-SRO changes the material behaviour and whether these changes can be accounted 
for using the developed continuum model. To the best of author's knowledge, there were
so far no studies reporting MD simulations of nanoindentation in CSA with C-SRO. 
After such simulations are carried out, the surface topographies should be 
rigorously studied. 
However, we expect that the combined model will be able to capture the mechanical
behaviour affected by the level of disorder by simply changing the value of a single
parameter $\alpha$. 

Other possible extensions or modifications of the model are straightforward, as
in other constitutive descriptions of plasticity. Note that it is relatively easy
to extend the model to address such phenomena as irradiation hardening
(cf. e. g. \cite{Han2012}) or indentation size effect (through strain gradient 
plasticity, cf. e. g. \cite{Rys2022}), especially exploiting the automatic 
differentiation and automatic code generation of modern software such as MFRONT
\cite{helfer_introducing_2015} or AceGen \cite{Korelc2002}. 
To the best of our knowledge, the effect of irradiation on the chemical disorder 
in CSAs was not thoroughly studied thus far. However, if the irradiation changes
the chemical disorder considerably, we expect that the single parameter $\alpha$ 
is going to capture this.

It is obvious that the proposed model accounts for phenomena emerging at the
atomistic level. In particular, both the indenter's radius and maximum penetration
depth were set to 5nm. It is now interesting to tell whether the effects observed
in MD simulations will be still present at the larger scale. 
The presented isotropic-crystal plasticity model was built in order to address the
effects observed in MD. At the same time it is scale-insensitive -- the same result
will be shown for tip radius at the scale of nano-, micro- or millimeters provided
that the depth-to-radius ratio is fixed. 
On the other hand, the observed "isotropization" of the surface topography is related
to dislocation patterns driven by atomic-scale chemical disorder. It should be
examined using dedicated experiments or simulations on larger material volumes
(e. g. discrete dislocation dynamics) whether this effect disappears when the
indented volume is larger. After answering this question, it will become clear
whether the proposed model is general and can be applied at a larger scale or
its area of application is limited to the nanometer scale. Moreover, if the 
effect of chemical disorder gradually diminishes with increasing the sample
volume, this will give rise to the size effect different from the one driven
by presence of geometrically necessary dislocations. However, the former one 
can be simply captured by varying the single parameter $\alpha$ and only the
treatment of the latter needs a strain gradient plasticity approach as e.g. in \cite{Rys2022}.